\documentclass[prl,twocolumn,aps,epsf,showpacs,superscriptaddress,shortbibliography]{revtex4-1}

\usepackage{amsmath}
\usepackage{amssymb}
\usepackage{graphicx}
\usepackage{hyperref}
\usepackage{float}
\usepackage{wasysym}
\usepackage{bbm}

\usepackage{titlesec}
\titleformat{\paragraph}[runin]% runin puts it in the same paragraph
        {\bfseries}% formatting commands to apply to the whole heading
        {}% the label and number
        {0.0em}% space between label/number and subsection title
        {}% formatting commands applied just to subsection title
        [~ -- ~]% punctuation or other commands following subsection title
\titlespacing*{\paragraph}{0pt}{4pt}{0pt}

\newcommand{\ket}[1]{|#1 \rangle}
\newcommand{\bra}[1]{\langle#1|}
\newcommand{\up}{\uparrow}
\newcommand{\down}{\downarrow}

\renewcommand{\(}{\left(}
\renewcommand{\)}{\right)}

\renewcommand{\]}{\right]}
\newcommand{\bs}[1]{\boldsymbol{#1}}

\newcommand{\N}{\mathcal{N}}

\hypersetup{colorlinks,citecolor=blue,linkcolor=blue,urlcolor=blue}

\newcommand{\beginsupplement}{%
        \setcounter{table}{0}
        \renewcommand{\thetable}{S\arabic{table}}%
        \setcounter{equation}{0}
        \renewcommand{\theequation}{S\arabic{equation}}%
        \setcounter{figure}{0}
        \renewcommand{\thefigure}{S\arabic{figure}}%
     }

\begin{document}

\title{Holographic dynamics simulations with a trapped ion quantum computer}
\author{Eli Chertkov}
\affiliation{Honeywell Quantum Solutions}
\affiliation{Institute for Condensed Matter Theory and IQUIST and Department of Physics, University of Illinois at Urbana-Champaign, Urbana, Illinois 61801, USA}
\author{Justin Bohnet}
\affiliation{Honeywell Quantum Solutions}
\author{David Francois}
\affiliation{Honeywell Quantum Solutions}
\author{John Gaebler}
\affiliation{Honeywell Quantum Solutions}
\author{Dan Gresh}
\affiliation{Honeywell Quantum Solutions}
\author{Aaron Hankin}
\affiliation{Honeywell Quantum Solutions}
\author{Kenny Lee}
\affiliation{Honeywell Quantum Solutions}
\author{Ra'anan Tobey}
\affiliation{Honeywell Quantum Solutions}
\author{David Hayes}
\affiliation{Honeywell Quantum Solutions}
\author{Brian Neyenhuis}
\affiliation{Honeywell Quantum Solutions}
\author{Russell Stutz}
\affiliation{Honeywell Quantum Solutions}
\author{Andrew C. Potter}
\affiliation{Department of Physics, University of Texas at Austin, Austin, TX 78712, USA}
\author{Michael Foss-Feig}
\affiliation{Honeywell Quantum Solutions}

\date{\today}

\begin{abstract}
Quantum computers have the potential to efficiently simulate the dynamics of many interacting quantum particles, a classically intractable task of central importance to fields ranging from chemistry to high-energy physics. However, precision and memory limitations of existing hardware severely limit the size and complexity of models that can be simulated with conventional methods. Here, we demonstrate and benchmark a new scalable quantum simulation paradigm---holographic quantum dynamics simulation---which uses efficient quantum data compression afforded by quantum tensor networks along with opportunistic mid-circuit measurement and qubit reuse to simulate physical systems that have far more quantum degrees of freedom than can be captured by the available number of qubits. Using a Honeywell trapped ion quantum processor, we simulate the non-integrable (chaotic) dynamics of the self-dual kicked Ising model starting from an entangled state of 32 spins using at most $9$ trapped ion qubits, obtaining excellent quantitative agreement when benchmarking against dynamics computed directly in the thermodynamic limit via recently developed exact analytical techniques.  These results suggest that quantum tensor network methods, together with state-of-the-art quantum processor capabilities, enable a viable path to practical quantum advantage in the near term.
\end{abstract}

\maketitle

Simulating the dynamics of many interacting quantum systems is a foundational problem in quantum science, underlying the computation of electronic and optical characteristics of materials and microelectronic devices, the prediction of chemical reaction kinetics, and even shedding light on the development of the early universe. Early exploration of quantum dynamics has already yielded fundamental insight into the quantum underpinnings of thermodynamics  and quantum chaos \cite{doi:10.1080/00018732.2016.1198134}  (and their alternatives~\cite{RevModPhys.91.021001}), and uncovered striking classes of universal behavior and critical phenomena in the structure of quantum many-body entanglement. Unfortunately, simulating quantum dynamics with classical computers is also a notoriously difficult problem, generically requiring resources scaling exponentially in either the size or evolution time of the simulated system. By contrast, it has been known for some time that quantum computers can simulate quantum dynamics with resources (qubit number and circuit depth) scaling only  polynomially \cite{Lloyd1073}.  For this reason, quantum dynamics simulation is widely regarded as a likely candidate for the first realization of \emph{practical} quantum advantage \cite{Childs9456}.

Large scale quantum simulations of simple models have been achieved in special-purpose quantum simulation platforms~\cite{Gross2017,Monroe2021,Browaeys2020}, but modeling realistic materials and processes will require universal, fully programmable quantum computers.  At present, however, such computers have small quantum memories (qubit numbers);  even as they surpass the scale of $\sim 50$ qubits necessary to provide a quantum advantage \emph{in principle}, much larger systems will be needed to simulate systems of typical physical interest, such as complex molecules or bulk materials.  Recently, a variety of quantum simulation algorithms based on quantum tensor-network (qTN) methods have been developed \cite{kim2017,Kim_2017,fossfeig2020,Liu_2019,barratt2020parallel,PRXQuantum.2.010342,smith2019crossing,yuan2020quantum,eddins2021doubling} that afford significant resource savings when simulating systems with less than maximal entanglement. Especially when combined with opportunistic mid-circuit measurements and reuse (MCMR) of qubits during a quantum computation~\cite{fossfeig2020}, qTN algorithms enable the simulation of systems with far more quantum degrees of freedom than can be directly mapped onto the qubits available in hardware.  When implemented on quantum computers with enough qubits to prohibit classical simulation, such algorithms may open an immediate path to quantum advantage in large-scale simulations of complex chemicals and materials.

In this work, we use Honeywell's H1 trapped ion quantum processor to implement the first experimental demonstration of the tensor-network-inspired \emph{holographic quantum dynamics simulation} (holoQUADS) algorithm \cite{fossfeig2020}, which we use to simulate an initially entangled 32-spin system evolving under chaotic quantum dynamics. Our results demonstrate that qTN methods, coupled with fidelities and technical capabilities currently available in state-of-the-art quantum computing platforms, afford exceptional quantitative accuracy in the simulation of quantum dynamics of large-scale many-body systems.

\begin{figure*}[t]
\begin{center}
\includegraphics[width = 0.9\textwidth]{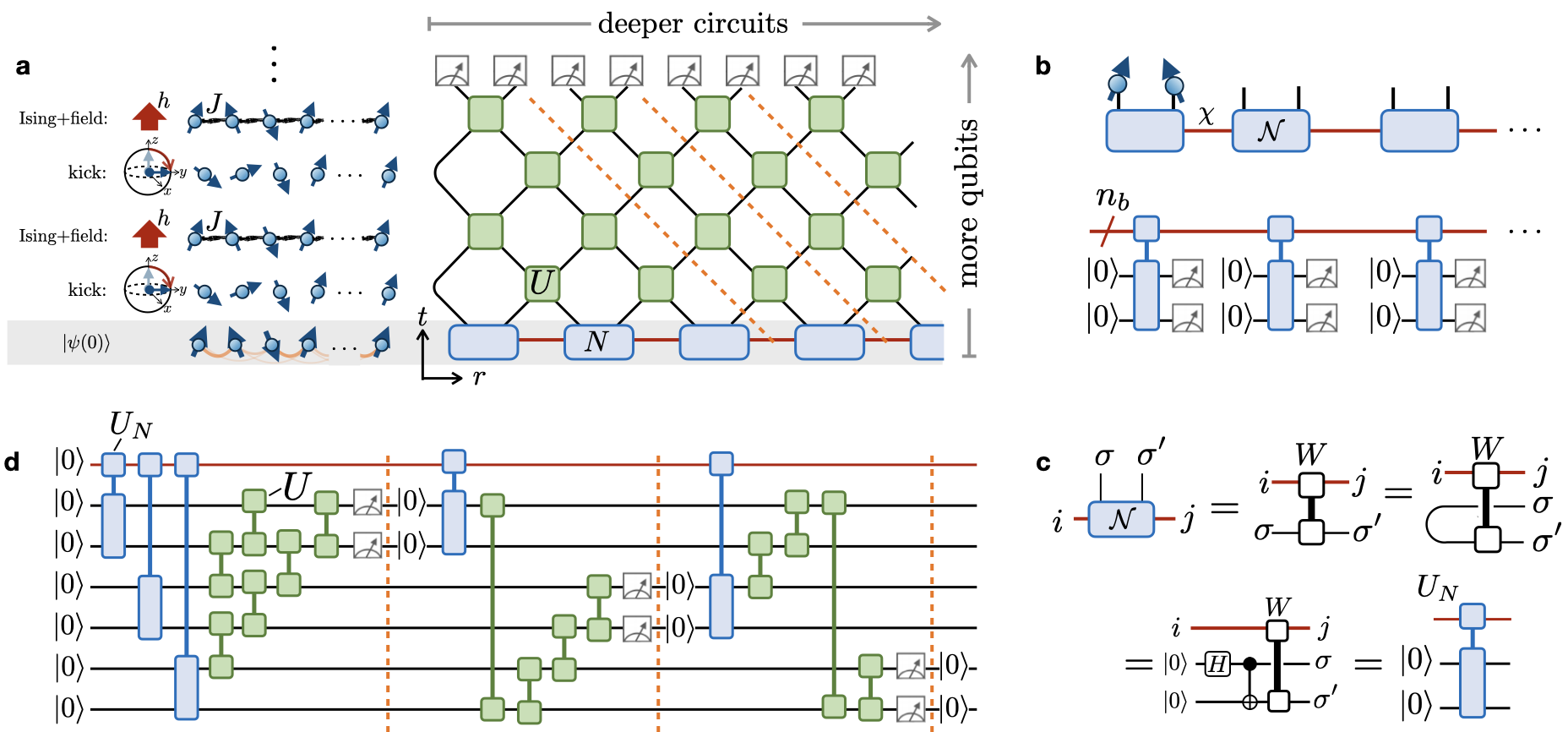}
\end{center}
\vspace{-0.25in}
\caption{{\bf Holographic simulation of the kicked Ising model.} \textbf{a} Schematic of the SDKI evolution 
($t=2$ Floquet periods shown) and its quantum circuit implementation ($t=4$ shown), 
which stroboscopically alternates between Ising interactions plus $z$-field and $\pi/2$-``kicks" by a transverse (x-axis) field. Green squares are dual-unitary gates, and blue rectangles represent site tensors for the initial correlated matrix product state, $\ket{\psi_0}$. \textbf{b}, \textbf{c} $\ket{\psi_0}$ is prepared holographically as a qMPS implemented by unitary circuits and MCMR as shown (with this example showing measurements of the qMPS in the $\sigma^z$ basis). For holoQUADS, measurements are postponed until after time-evolution as shown in \textbf{a}. \textbf{d} The same state-preparation and time-evolution redrawn as a holographic quantum circuit using mid-circuit measurements and qubit recycling (the holoQUADS algorithm). 
}\label{fig:circuits}
\end{figure*}

With conventional Hamiltonian simulation techniques, simulating the dynamics of a system of size $L$ requires $O(L)$ qubits, and the achievable simulation time $t$ is limited by gate fidelities. By interchanging space (qubit number) and time (circuit depth) resource scaling, holoQUADS enables time-evolution of arbitrarily large systems up to a time $t$ set by the number of available qubits, independent of $L$, thereby directly targeting the typical ``thermodynamic'' limit ($L \rightarrow \infty$) of interest to physicists. Moreover, holoQUADS leverages a compact representation of dynamics starting from initial correlated states, such  as low-energy states of many-body Hamiltonians, requiring only (roughly) one additional qubit per unit of bipartite entanglement in the initial state. By contrast, traditional techniques require complicated circuits to prepare such states either through adiabatic state preparation or variational means. 

Taken together, these features ensure that qubit resources are not only much lower than with conventional methods, but also that they are optimally utilized: Time evolution of a state with initial entanglement entropy $S_0$ generically produces entanglement bounded by $S(t)\lesssim S_0+c t$ (with $c$ a model-dependent constant) \cite{calabrese2005evolution}. Since holoQUADS requires one qubit per bit of initial entanglement plus an additional number of qubits linear in $t$ to enact the time evolution, resulting in a total qubit scaling $N\sim S(t)$, we see that every qubit is allocated directly towards the classically hard feature of time evolution in many-body quantum systems: the growth of entanglement.

\paragraph{HoloQUADS algorithm}
HoloQUADS~\cite{fossfeig2020} simulates the time-evolution of correlated initial states expressed as matrix product states (MPS).  For this work, we consider MPS of a half-infinite system with translationally invariant tensors over a two-site unit cell, which can be written as
\begin{align}
	\ket{\psi_0} =\!\!\!\!\!\!\!\! \sum_{\sigma_1,\sigma_2,\dots\in \{\up,\down\}} \!\!\!\!\!\!\!\! \ell^T\mathcal{N}^{(\sigma_1,\sigma_2)}\mathcal{N}^{(\sigma_3,\sigma_4)} \cdots  \ket{\sigma_1\sigma_2\sigma_3\sigma_4 \cdots }.\label{eq:solvableMPS}
\end{align}
Here $\ket{\sigma_1\sigma_2\cdots}$ is the eigenstate of all Pauli operators $\sigma^z_i$ with eigenvalues $\sigma_i$, $\mathcal{N}$ are rank-4 tensors, i.e., $\chi\times \chi$ matrices for each spin configuration $(\sigma_i,\sigma_{i+1})$, and the bond-dimension $\chi$ controls the amount of entanglement present in the state (very roughly, to capture a state with bipartite entanglement $S$, one needs $\chi\sim e^S$).
As originally discussed in Ref.~\cite{Schon_2005}, any MPS can be realized as a quantum circuit where the tensors $\mathcal{N}$ are implemented by a unitary $U_{\N}$ acting on a pair of ``system" qubits (representing a pair of neighboring spins) initialized to $\ket{0}$, and $n_b=\lceil \log_2 \chi \rceil$ ``bond" qubits that represent the $\chi$-dimensional bond space.  Following the language in \cite{barratt2020parallel}, we refer to MPS prepared this way on a quantum processor as quantum MPS (qMPS).

While a length $2L$ section of the qMPS described above naively requires $n_b + 2L$ qubits to represent, MCMR allows us to ``holographically'' represent it using only $n_b+2$ qubits \cite{Huggins_2019,fossfeig2020}, as shown in Fig.\,\ref{fig:circuits}\textbf{b} \footnote{This construction of a qMPS is ``holographic'' in the sense that a $1$D system is being simulated using a $0$D system (constant number of qubits).}. An MPS evolving in time under the influence of a $1$D quantum circuit with layers of gates acting on neighboring qubits also admits a holographic representation (see Fig.\,\ref{fig:circuits}) \footnote{This mapping does compromise some (but not all) of the 1D character of the time-evolution circuit.  In the QCCD architecture \cite{wineland1998experimental,kielpinski2002,pino2020}, where arbitrary rearrangement of qubits is possible prior to gating, this induced non-local character incurs very little cost.}. Due to the causal ``light cone" structure of the quantum circuit and the qMPS initial state, the measurement results at the top (larger $t$) of the circuit are only affected by a subset of qubits at the bottom (smaller $t$). The circuit can therefore be executed going from left to right in slices bounded by orange dashed lines in Figs.\,\ref{fig:circuits}\textbf{a},\textbf{d}, where in each slice two qubits are reset for later reuse.  Simulating $t$ layers of nearest-neighbor time-evolution of the $\chi$ bond-dimension MPS in Eq.\,(\ref{eq:solvableMPS}) with holoQUADS requires $n_b+t+2$ qubits  ($n_b=\lceil \log_2 \chi \rceil$) \footnote{Generally, holoQUADS can simulate an MPS (with a single-qubit unit cell) time-evolved by $t$ layers of a $1$D circuit with $k$-local gates using $n_b + t + k-1$ qubits \cite{fossfeig2020}.}, giving a logarithmic resource reduction compared to classical TN techniques (which require memory scaling polynomially with $\chi$ and exponentially in $t$). Note that the scaling is independent of system size $L$ \footnote{For non-critical states the number of bond qubits required to achieve fixed accuracy is independent of $L$; at criticality there would be $\sim\log L$ corrections to the qubit requirements.}, so holoQUADS can be used to time-evolve an arbitrarily large qMPS.

 \begin{figure}[t]
\begin{center}
\includegraphics[width = 1.0\columnwidth]{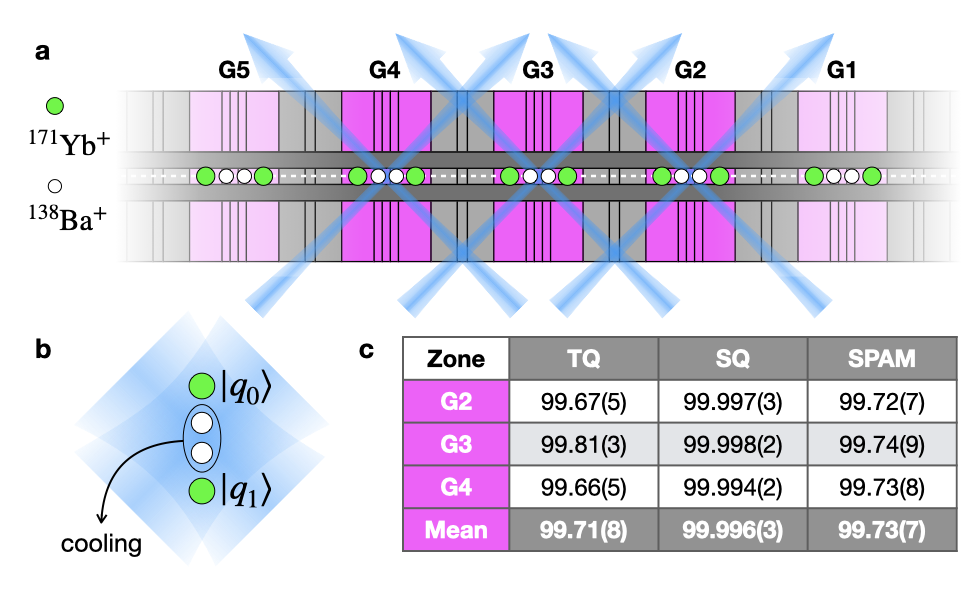}
\end{center}
\vspace{-0.25in}
\caption{{\bf Quantum computer used in this work.} \textbf{a} Section of the Honeywell H1 segmented-electrode surface trap, showing 5 gate zones in purple (each is $750\,\mu{\rm m}$ wide, ions crystal extents and laser beam waists are not drawn to scale).  The computer operates similarly to the one described in Ref.~\cite{pino2020} [except with parallel gate operation across the central three gate zones (G2-G4)], with $^{171}{\rm Yb}^+$ qubit ions (green) and $^{138}{\rm Ba}^+$ coolant ions (white) stored in either 2-ion or 4-ion crystals. Arbitrary pairing of qubits is achieved by transporting ions \cite{PhysRevA.90.053408,10.5555/2012086.2012087,Splatt_2009} along the linear RF-null (dashed line) $70 \mu$m above the surface.  \textbf{b} Sympathetic ground state cooling followed by our two-qubit, phase-insensitive M\o lmer-S\o renson gate is implemented in parallel across G2-G4 on Yb-Ba-Ba-Yb crystal configurations. Each crystal is roughly $8\,\mu{\rm m}$ in extent, and the cooling and gate lasers (at wavelengths of $493\,$nm and $368\,$nm, respectively) have nominal beam waists of $17.5\,\mu{\rm m}$  \textbf{c} Typical (i.e., representative over the duration of data taking) average fidelities of single-qubit (SQ), two-qubit (TQ), and combined state-preparation and measurement (SPAM), measured via randomized benchmarking.}
\label{fig:architecture}
\end{figure}

\paragraph{A chaotic circuit dynamics benchmark}
As a first demonstration of holoQUADS, we focus on the dynamics of the self-dual kicked Ising (SDKI) model~\cite{Akila2016,bertini2018}. In this model, a spin chain evolves under an Ising interaction and an integrability-breaking longitudinal field $h$ and is periodically ``kicked'' by a transverse (X) field:
\begin{align}
H(t) = \sum_{i} \(\frac{\pi}{4} \sigma^z_i\sigma^z_{i+1}+h \sigma^z_i + \frac{\pi}{4} \sum_{n\in \mathbb{Z}}\delta(t-n) \sigma^x_i\).
\end{align}
Here $\bs{\sigma}_i$ are Pauli operators on site $i$ of a length $2L$ spin chain.  For $h\neq 0$ the SDKI is non-integrable~\cite{bertini2019}, exhibiting the chaotic and thermalizing behavior expected generically for non-integrable quantum dynamics.  Moreover, recent analytical techniques enable exact calculations of many of its properties \cite{bertini2018,bertini2019,PhysRevB.100.064309}, rendering it a powerful benchmark for the performance of holoQUADS.

\begin{figure*}[t]
\begin{center}
\includegraphics[width = 1.0\textwidth]{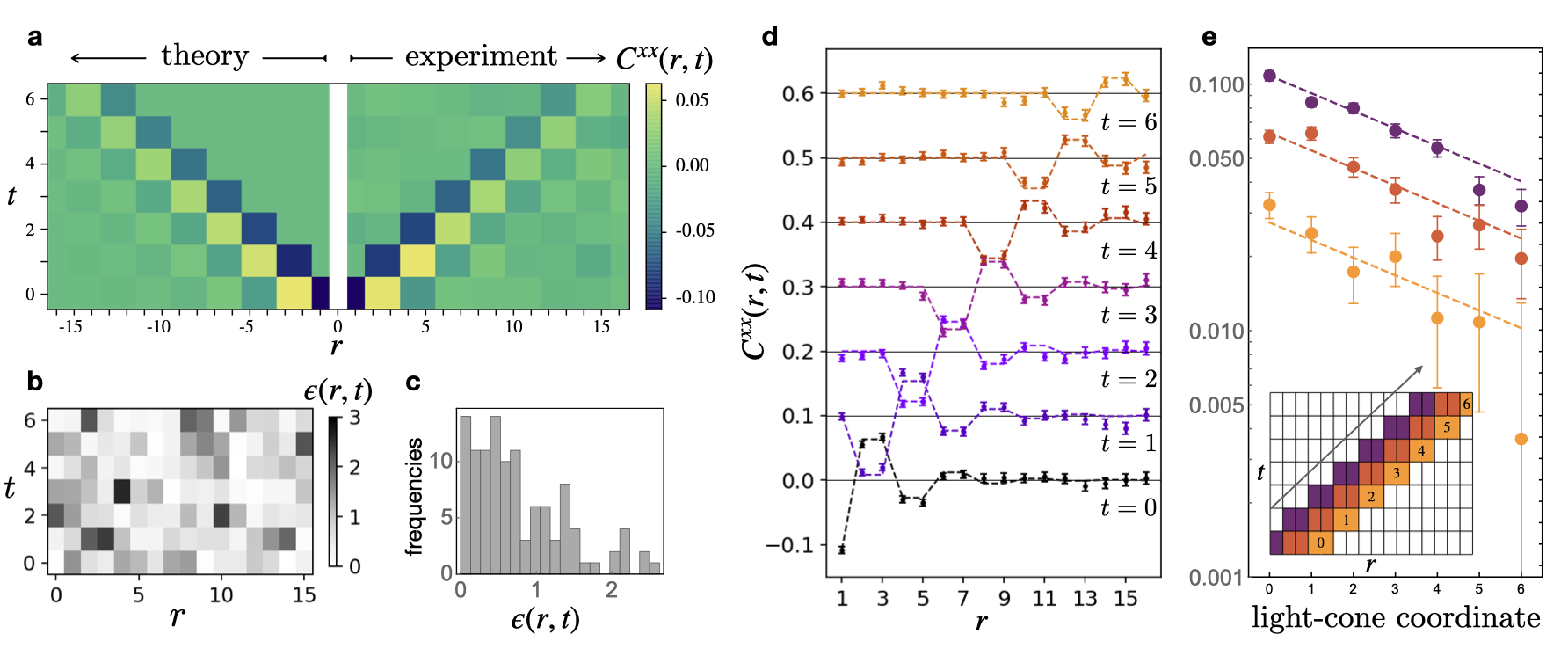}
\end{center}
\vspace{-0.25in}
\caption{{\bf Experimental data.} \textbf{a} Color plot of equal-time spin-spin correlators $C^{xx}(r,t)$ for the SDKI model with $h=0.2$ starting from a qMPS with the parameters $(K_x,K_y,K_z)=(0.3,0.5,1.25)$ explored in Ref.~\cite{piroli2020}, showing theoretical results (left side, $r<0$) and experimental results (right, $r>0$). The correlators exhibit two characteristic features of generic dual-unitary circuit dynamics: correlations spread with maximal velocity along ``light cones" of the circuit and correlations decay exponentially along the light cones. Note that the data in this plot are aggregated into bins ($r\in\{2j,2j+1\}$ for $j>0$) containing symmetry equivalent sites in order to smooth and reduce statistical fluctuations. \textbf{b} Gray-scale heatmap and \textbf{c} histogram of $\epsilon(r,t)$, the absolute difference between experimental and theoretical correlation functions normalized by the standard error (due to finite sampling) of the experimentally measured $C^{xx}(r,t)$ values [see Eq.~(\ref{eq:error})]. \textbf{d} Traces of $C^{xx}(r,t)$, with data offset vertically by $0.2t$ for clarity, with dots showing experimental data, error bars showing standard error due to finite sampling, and dashed lines showing theoretical results in the thermodynamic limit ($L\rightarrow \infty$). \textbf{e} Log-linear plot of correlations along three different light cone trajectories (the precise set of $r$ and $t$ values for each curve are indicated in the inset), with different offsets, $k=0,1,2$ from the center of the light cone. Each shows characteristic exponential decay of correlations along the light cone.
}\label{fig:expt}
\end{figure*}

As illustrated in Fig.\,\ref{fig:circuits}\textbf{a}, the time-evolution operator of the SDKI model, $\mathcal{U}_t = \mathcal{T}\{e^{-i\int_0^t H(s)ds}\}$, can be recast (at discrete times $t\in \mathbb{Z}$ and up to boundary terms in time) as a 1D quantum circuit consisting of alternating even-odd layers of two-qubit gates \cite{PhysRevB.100.064309,bertini2019}
\begin{align}
U &= (u_+ \otimes u_-) e^{-i\frac{\pi}{4}\left(\sigma^x \otimes \sigma^x + \sigma^y \otimes \sigma^y \right)}(v_- \otimes v_+).
\end{align}
Here, $u_+ = e^{-ih\sigma^z}e^{i\frac{\pi}{4}\sigma^x}e^{-i\frac{\pi}{4}\sigma^y}$, $u_- = e^{i\frac{\pi}{4}\sigma^x}e^{-i\frac{\pi}{4}\sigma^y}, v_- = e^{i\frac{\pi}{4}\sigma^y}e^{-ih\sigma^z}$, and $v_+=e^{i\frac{\pi}{4}\sigma^y}$ are single-qubit rotations. For any single-qubit unitaries $u_\pm,v_\pm$, this gate satisfies a special dual-unitary property~\cite{PhysRevB.100.064309,bertini2019}: $U^\dagger U^{\vphantom\dagger}= \mathbbm{1} = \tilde{U}^{\dagger}\tilde{U}^{\vphantom\dagger}$ where the dual $\tilde{U}$ of a two-qubit gate $U$ is defined by a reshuffling of indices, $\bra{k} \otimes \bra{l} \tilde{U} \ket{i} \otimes \ket{j} = \bra{j} \otimes \bra{l} U \ket{i} \otimes \ket{k}$.

Dual-unitary circuits can be interpreted as circuits that generate time-evolution both in time \emph{and space}.  Many of their properties, such as spectral form factors and entanglement spectra, spread of local operator correlations, and out-of-time-ordered correlators can be computed analytically for special initial conditions: either infinite temperature states \cite{bertini2019} or ``exactly-solvable" matrix product states (MPS) \cite{piroli2020} of the form in Eq.\,(\ref{eq:solvableMPS}) with
\begin{align}
\N^{(\sigma,\sigma')}_{i,j} = \bra{j} \otimes \bra{\sigma'} W \ket{i} \otimes \ket{\sigma}
\label{eq:W}
\end{align}
defined by a unitary matrix $W\in U(2\chi)$ \footnote{Strictly speaking, the initial state should satisfy Eq.\,(\ref{eq:W}) \emph{and} the MPS should be injective \cite{piroli2020}}.  While the dual-unitary property is clearly fine-tuned and results in some peculiar features that do not survive generic perturbations, such as correlations spreading at the maximum possible velocity, dual-unitary circuits typically (e.g., for $h\neq 0$ in the SDKI model) are non-integrable and exhibit  generic features of chaos, ergodicity, and thermalization~\cite{bertini2020,claeys2021}. This combination of striking features and solvability make the SDKI model an important minimal model of quantum dynamics and, for our purposes, a useful benchmark for dynamical simulation.  Reference ~\cite{piroli2020} derived exact thermodynamic-limit expressions for equal-time two-point correlation functions of Pauli operators. Following that work, we consider the smoothed correlation functions
\begin{align}
C^{\alpha\beta}(r,t) = \frac{1}{4L}\sum_{j=1}^{2L} \sum_{\delta=0,1} \langle\psi_t|\sigma^\alpha_{j}\sigma^\beta_{j+r+\delta}|\psi_t\rangle, \label{eq:Cab0}
\end{align}
for which the sum over $\delta$ removes an even/odd effect, making the correlations easier to visually interpret.  Here $\ket{\psi_t}$ is the wave function after the $t^{\rm th}$ layer of the circuit, as shown in Fig.\,\ref{fig:circuits}\textbf{a}.

\paragraph{Implementation on a trapped ion quantum processor}
We implement holoQUADS of the SDKI model for $2L=32$ spins on Honeywell's H1 trapped ion quantum processor (see Fig.\,\ref{fig:architecture}), using between 3-9 $^{171}\text{Yb}^+$ hyperfine qubits, depending on the number of simulated circuit layers (see Table~\ref{tab:experiment_parameters}). Note that there is no strict limitation on $L$, since holoQUADS (as implemented here) simulates a half-infinite chain, and larger $L$ would simply have incurred marginally longer run times on the quantum computer (scaling linearly with $L$).  However, correlations outside the 32-site measurement window are negligible for the initial correlation length and evolution time accessible with currently available qubit numbers.  

The MCMR operations that enable holographic algorithms are performed with high-fidelity by temporarily separating \cite{Barrett:2004vw,Chiaverini:2004uj} the targeted and spectator ions from one another by a distance that is large ($\gtrsim 180 \mu$m) compared to the $1/e^2$ radius of the resonant laser beam (between $13 \mu$m and $20 \mu$m depending on the zone) used for measurement and reset~\cite{Olmschenk2007}. Cross-talk is further suppressed by temporarily moving spectator ions off the RF-null, which Doppler shifts the light from resonance~\cite{Berkeland1998}, resulting in cross-talk errors on spectator qubits of $\lesssim 1\times 10^{-3}$ for resets and $\lesssim 5\times 10^{-3}$ for measurements in the worst case (i.e., for a spectator qubit in the worst possible location), and nearly an order of magnitude lower than that on average.

\begin{table}
\begin{center}
\begin{tabular}{|c|c|c|c|c|c|c|c|}
\cline{2-8}
\multicolumn{1}{c|}{} & $t=0$ & $t=1$ & $t=2$ & $t=3$ & $t=4$ & $t=5$ & $t=6$ \\
\hline
\# of qubits & 3 &5 &5 &7 &7 &9 &9 \\ \hline
\# of SQ gates & 214 &273 &308 &360 &394 &454 &488 \\ \hline
\# of TQ gates & 66 &87 &104 &129 &146 &175 &192 \\ \hline
$\%$ leaked &  $3.1$ & $1.7$ & $2.0$ & $1.9$ & $2.4$ & $3.6$ & $3.7$ \\ 
\hline
\end{tabular}
\end{center}
\caption{{\bf Experimental resources.} The required resources and detected leakage for the seven dual-unitary holoQUADS experiments. Each column corresponds to simulated time-evolution of a half-infinite matrix product state for time $t$.  For each value of $t$ the associated circuit extracts correlation functions over 32 lattice sites, which involved 32 mid-circuit measurements and qubit resets, and each circuit was repeated 1000 times to gather statistics. The first row lists how many qubits were used in each experiment. The second and third rows list how many single-qubit (SQ) and native two-qubit (TQ) gates were used. The last row indicates the percentage of the $1000$ experimental trials that were discarded due to detected leakage of the bond qubit (see appendix).} 
\label{tab:experiment_parameters}
\end{table}

We prepare an initial qMPS corresponding to Eq.\,(\ref{eq:W}) with $W=\exp[-i\sum_{\alpha=x,y,z}K_\alpha \sigma^\alpha \otimes \sigma^\alpha]$. Using the circuit identities shown in Fig.\,\ref{fig:circuits}\textbf{c}, we can implement the tensors $\mathcal{N}$ (corresponding to $W$) as a unitary circuit by creating a Bell pair of the physical qubits to appropriately reorder the qubit lines. We ran seven holoQUADS time-evolution circuits (one for each duration $t=0,1,..,6$ of time-evolution), in each case executing $16$ ``slices'' of holoQUADS (see Fig.\,\ref{fig:circuits}), resulting in a simulation of a 32-site MPS. The resource requirements for these circuits, such as the number of two-qubit gates used, are summarized in Table~\ref{tab:experiment_parameters}. Each circuit was repeated 1000 times in order to reconstruct estimates of the correlation function $C^{xx}(r,t)$~\footnote{To avoid boundary effects when computing 
$C^{xx}(r,t)$ in Eq.~(\ref{eq:Cab0}), we averaged over positions from $j=9$ to 
$j=2[L - \text{ceiling}(r/2)-1]$%\left\lceil r/2 \right\rceil - 1)$ 
instead of averaging over all $j=1,\ldots,2L$.  Note that in a single experiment we gathered data for multiple $r$ values in $C^{xx}(r,t)$.
}. The experimental results are summarized in Fig.\,\ref{fig:expt}, and show excellent quantitative agreement with exact theoretical results in the thermodynamic limit (dashed lines in Fig.\,\ref{fig:expt}\textbf{d}). We clearly observe~\footnote{For the light cone plot in Fig.\,\ref{fig:expt}\textbf{a}, we exploit symmetry based arguments to aggregate experimental data into pairs of even-odd-sites that must share the same values of the smoothed correlation function defined in Eq.\,(\ref{eq:Cab0})} 
that information propagates with maximal velocity along a sharp light cone (Fig.\,\ref{fig:expt}\textbf{a}), which is the hallmark of dual-unitary circuit dynamics \cite{claeys2020}, and that correlations decay exponentially along the light cone, indicating the ergodic (non-integrable) character of the dynamics (Fig.\,\ref{fig:expt}\textbf{e}).  Note that the data is almost entirely unprocessed. The only form of error mitigation we apply is to detect leakage of the bond qubit out of the qubit-state manifold at the end of the holoQUADS algorithm (see appendix). The results are post-selected on experimental trials without bond-qubit leakage, which amounts to neglecting less than $3\%$ of the total data (see Table~\ref{tab:experiment_parameters} for leakage statistics).

To better isolate statistical (i.e., finite sampling) errors from intrinsic errors due to circuit noise, imperfect qubit control, and finite-size effects, we calculate the normalized errors:
\begin{align}
	\epsilon(r,t) = \left\vert C^{xx}_\text{expt}(r,t) - C_{\infty}^{xx}(r,t)  \right\vert/\Delta C^{xx}(r,t). \label{eq:error}
\end{align}
Here $C^{xx}(r,t)_\text{expt}$ is the experimentally estimated correlation function, $ C^{xx}_{\infty}(r,t)$ is the exact theoretical result in the thermodynamic limit, and 
$\Delta C^{xx}(r,t) = (\text{Var}\[C^{xx}(r,t)\]_\text{expt}/N_{\rm s})^{1/2}$ is the standard error.
We find that the normalized errors $\epsilon(r,t)$ differ from the theoretical results by $0.77$ standard errors on average over all $(r,t)$ (see Fig.\,\ref{fig:expt}\textbf{b},\textbf{c}). In this sense, at the level possible given statistical errors from finite size sampling, the quantum computation is providing quantitively correct predictions of the non-equilibrium dynamics of a large, initially correlated quantum spin system.

\paragraph{Discussion}

Our results showcase both the viability of quantum computers for solving classically hard and practically relevant models of many-body quantum dynamics, and the benefits of MCMR for simulating complex quantum dynamics of large, highly correlated quantum systems using quantum processors with limited qubit numbers. We emphasize that while we have demonstrated the holoQUADS algorithm for a finely-tuned dual-unitary circuit, we have done so for benchmarking purposes. Dual-unitary circuits admit a convenient classical short-cut to solution despite worst-case (from a classical simulation perspective) ballistic growth of entanglement, while retaining many of the features expected of generic quantum dynamics.  Our implementation of holoQUADS does not take advantage of any of the fine-tuned self-dual structure of these circuits, and can be implemented to achieve significant resource savings in any situation where one would like to time evolve a correlated but not maximally-entangled initial state.  

While our current results can readily be simulated by classical time-evolved block decimation (TEBD) methods (or by directly simulating the quantum algorithm, which in this work involved no more than $9$ qubits), the excellent agreement with exact results at all simulation times demonstrates that we are currently limited by our qubit numbers, rather than our gate fidelities.  As quantum hardware with comparable gate fidelities and larger qubit numbers becomes available, enabling deeper time evolutions of systems directly in the thermodynamic limit, we expect that qTN methods will enable further progress towards the ultimate goal of outperforming classical simulation capabilities on problems of direct physical relevance. Obvious targets for quantum advantaged dynamics simulation include long-time dynamics of $1$D systems for which the cost of classical TEBD methods grows exponentially in time due to linear-in-time entanglement growth, the dynamics of $2$D and $3$D systems which can be represented holographically as isometric tensor networks~\cite{zalatel2020}, or the dynamics of systems with longer-range (e.g., truncated Coulomb) interactions.

\emph{Acknowledgements.---} This work was made possible by a large group of people, and the authors would like to thank the entire Honeywell Quantum Solutions team for their many contributions.  We thank Charlie Baldwin for helpful discussions regarding performance benchmarks on H1.  We used the ITensor library \cite{itensor}, written in Julia \cite{julia}, to perform the tensor network contractions to generate the theory curves. Quantum circuits were prepared and simulated using the Qiskit library created by IBM \cite{qiskit}. ACP was supported by NSF Convergence Accelerator Track C award 2040549.

\bibliography{refs}

\appendix
\beginsupplement

\section{SWAP-replacement in quantum circuits}
\begin{figure}[H]
\begin{center}
\includegraphics[width=0.5\textwidth]{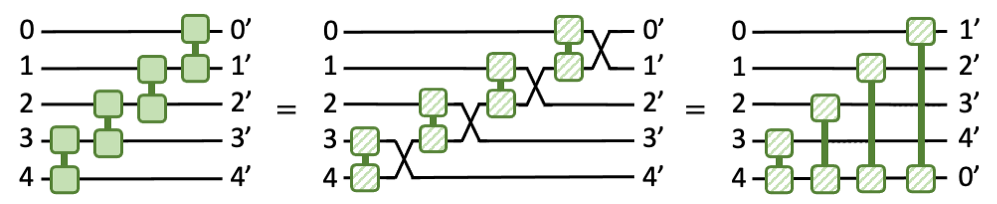}
\end{center}
\caption{A circuit identity for a chain of dual-unitary gates. SWAP gates can be factored out of the dual-unitary gates at the cost of introducing long-range gates and relabeling the qubits after the chain. For the SDKI model, each new (hatched) gate can be implemented with one fewer native two-qubit gate than the original (solid) gate.} \label{fig:swap_trick}
\end{figure}

The largest source of error in our circuits is imperfections in native two qubit gates on the Honeywell H1 quantum computer, which are a rotated version of the M\o lmer-S\o renson gate~\cite{Sorensen2000}
\begin{align}
\overline{\rm MS}=e^{i(\pi/4)\sigma^z \otimes \sigma^z}.
\label{eq:def_MS}
\end{align}
This entangling gate is equivalent to CZ or CNOT up to single-qubit gates, and operates at $99.71(8)\%$ fidelity averaged across the three gate zones used in this work. We can reduce the impact of such errors by using circuit identities, together with the ability to physically rearrange qubits in the QCCD architecture to directly execute long-range gates, in order to minimize the total two-qubit gate count.  Up to single-qubit rotations, any dual-unitary gate takes the form
\begin{align} 
V[J]&=\exp\left[-i\left(\frac{\pi}{4}\sigma^x \otimes \sigma^x + \frac{\pi}{4}\sigma^y \otimes \sigma^y + J\sigma^z \otimes \sigma^z\right)\right] 
\nonumber\\
&\propto e^{i(-J+\frac{\pi}{4})\sigma^z \otimes \sigma^z} \text{SWAP},
\end{align}
where the proportionality is up to an irrelevant overall phase. As shown in Fig.\,\ref{fig:swap_trick}, the $\text{SWAP}$ gates can be factored out of the dual-unitary gates in a slice of the holoQUADS algorithm. In the QCCD architecture, the SWAP gates can be implemented by physical ion transport (essentially error-free) without any error-prone logical operations (equivalently, we can permute qubit labels and redraw the circuit without SWAPs but with long-range gates). For the SDKI model that we consider, where $J=0$, the original dual-unitary gate can be represented with two native two-qubit gates, but the SWAP-factored gates can be represented with only one native gate. Therefore, this replacement roughly provides a factor of two savings in the number of two-qubit gates.

\section{Leakage detection}

\begin{figure}[!t]
\begin{center}
\includegraphics[width=0.8\columnwidth]{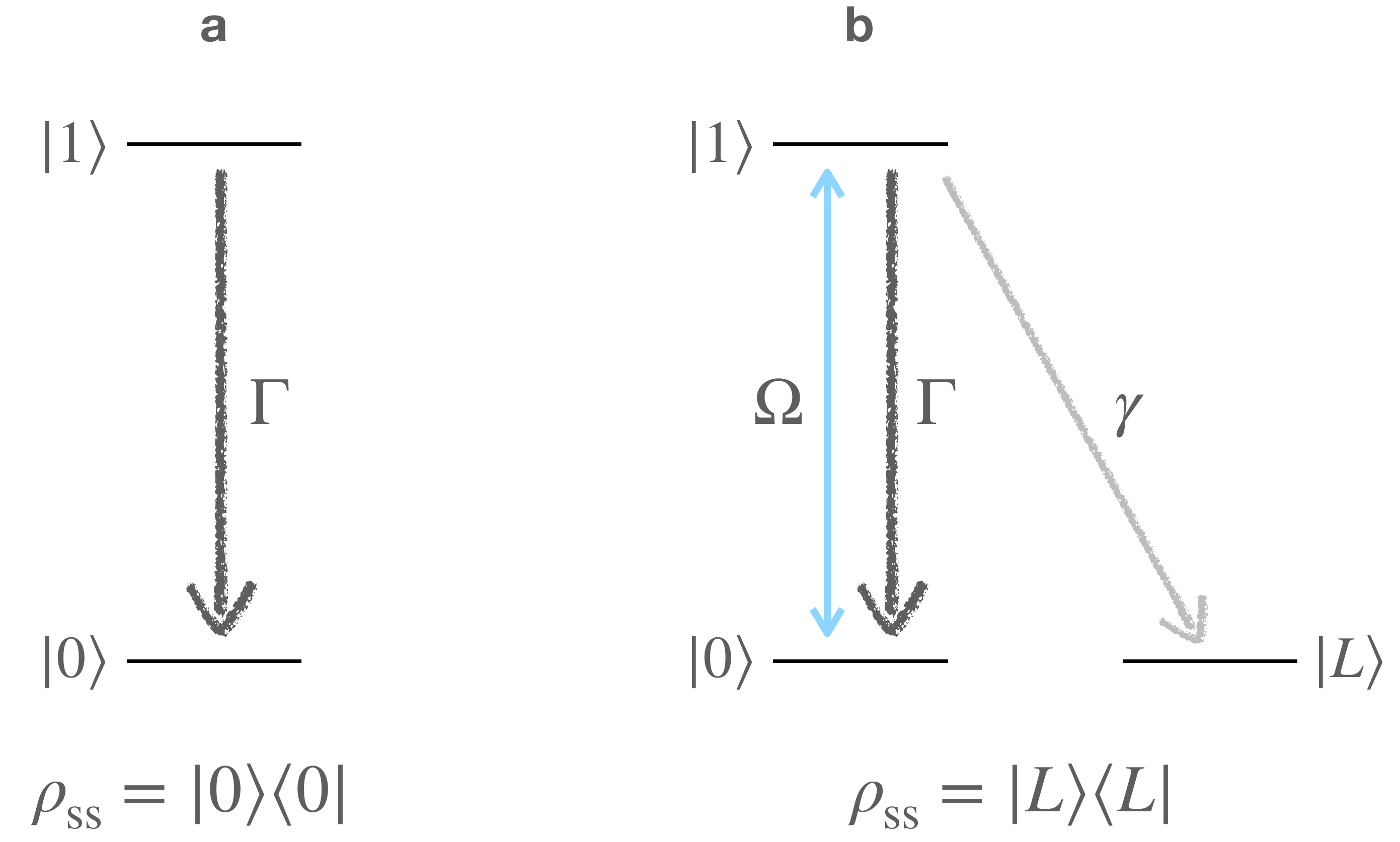}
\end{center}
\caption{An example of how a gapped steady state can be destroyed by arbitrarily small leakage.  {\bf a}  A two level system (qubit) with spontaneous emission from $\ket{1}\rightarrow\ket{0}$  has a unique and gapped steady state $\ket{0}\bra{0}$.  {\bf b} In the presence of arbitrarily weak coherent driving on the $\ket{0}\leftrightarrow\ket{1}$ transition and arbitrarily weak spontaneous emission to a leaked state $\ket{L}$, the steady state is completely modified to $\ket{L}\bra{L}$ (e.g., in optical pumping).    }
\label{fig:steady_state}
\end{figure}

Any MPS can be recast as a quantum channel defined on its bond-indices, and the qMPS preparation circuit can be viewed as a purification of that channel.  For an MPS with finite correlation length, the channel has a non-degenerate and gapped steady state, such that all observables converge to a unique value over a time-scale (length-scale) corresponding to the channel memory time (MPS correlation length).  Since the steady state of the channel is gapped, qMPS observables should be stable to (i.e., weakly perturbed by) small errors/perturbations to the gates generating the qMPS.  However, perturbations that cause leakage out of the qubit subspace can be singular, as the channel will not generally remain gapped when defined over the enlarged Hilbert space accessed by the leakage process.  As a simple example, consider spontaneous emission in a two-level system as shown in Fig.\,\ref{fig:steady_state}\textbf{a}, for which the steady state $\ket{0}\bra{0}$ is unique and the Liouvillian has gap $i\Gamma$.  In the presence of arbitrarily weak driving (rate $\Omega$) and arbitrarily weak spontaneous emission to a leaked state $\ket{L}$ (rate $\gamma$), the new steady state $\ket{L}\bra{L}$ is orthogonal to the steady state at $\Omega,\gamma=0$.

\begin{figure}[H]
\begin{center}
\includegraphics[width=0.3\textwidth]{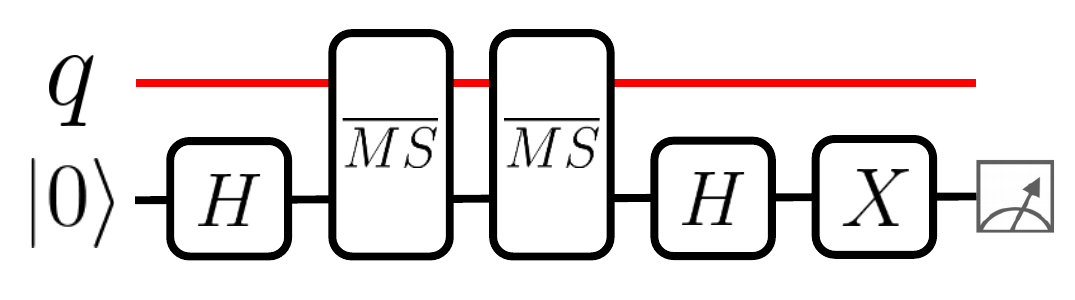}
\end{center}
\caption{The leakage detection gadget used to detect if qubit $q$ leaked. $\overline{\rm MS}$ is the rotated M\o lmer-S\o renson gate defined in Eq.\,(\ref{eq:def_MS}).}
\label{fig:leakage_detection_gadget}
\end{figure}

The primary sources of leakage errors in the H1 quantum computer are spontaneous emission during the $\overline{\rm MS}$ gate and measurement/reset cross talk.  The qubit states $\ket{0(1)}$ are chosen to be the $\ket{F=0(1),m_F=0}$ levels of the electronic $^2 S_{1/2}$ ground-state hyperfine manifold, and during two-qubit gates or measurement/reset of nearby qubits there is a small probability for a qubit to undergo spontaneous Raman scattering into the states $\ket{F=1,m_F=\pm 1}$. Most qubits in our circuits are occasionally reset for reuse, which fixes leakage errors.  The bond qubit, however, propagates through the entire circuit without reset, and so even very small leakage errors can eventually accumulate. At the system sizes and evolution times accessed here, this leakage has a small but nevertheless noticeable impact on circuit fidelities, and so we post select on data for which the bond-qubit has not leaked by using the circuit-level leakage detection protocol shown in Fig.\,\ref{fig:leakage_detection_gadget}.

When the $\overline{\rm MS}$ gate is applied between a leaked qubit and another qubit it has no effect (it acts as an identity operator) \cite{hayes2012}. This fact allows us to use two $\overline{\rm MS}$ gates and an ancilla qubit to directly measure if a qubit has leaked (see Fig.\,\ref{fig:leakage_detection_gadget}).  If neither qubit has leaked, two repeated applications of the $\overline{\rm MS}$ gate yields a global single-qubit $z$-rotation by $\pi$, i.e.\ $(\overline{\rm MS})^2=\sigma^z\otimes\sigma^z$ up to an irrelevant global phase. In this case the ancilla qubit ends up in $ \sigma^x H \sigma^z H\ket{0} = (\sigma^x)^2\ket{0} = \ket{0}$. If $q$ has leaked, then the ancilla qubit is mapped to $\sigma^x H I H \ket{0} = \sigma^x H^2 \ket{0}=\sigma^x\ket{0}=\ket{1}$, heralding the leakage event. Note that if leakage is not detected, then $\sigma^z$ is applied to qubit $q$; this could be undone by applying a classically conditioned $\sigma^z$ gate on qubit $q$ (applied when the ancilla measures a $\ket{0}$) if one desires a QND leakage measurement, e.g.\ in order to measure leakage in the middle of a circuit without affecting the state of $q$ in the event that it has not leaked.

\end{document}